\documentclass[onecolumn]{IEEEtran}
\pdfoutput=1
\usepackage{cite}
\usepackage{amsmath,amssymb,amsfonts}
\usepackage{algorithmic}
\usepackage{graphicx}
\usepackage{textcomp}
\usepackage{color}
\usepackage[capitalize]{cleveref}
\usepackage{soul}
\usepackage{balance}

\usepackage{epsfig}
\usepackage{amsfonts}
\usepackage{listings}
\usepackage[utf8]{inputenc}

\usepackage{lmodern}
\usepackage{mathtools}
\usepackage{graphicx}
\usepackage{textcomp}
\usepackage{subfigure}
\usepackage{epstopdf}
\usepackage{enumerate}
\usepackage{color}
\usepackage{cleveref}

\usepackage{stmaryrd}
\usepackage{epstopdf}

\newtheorem{theo}{Theorem}
\newtheorem{lem}{Lemma}

\newtheorem{rem}{Remark}

\newtheorem{defi}{Definition}
\newtheorem{example}{Example}
\newtheorem{assum}{Assumption}
\newtheorem{proof}{Proof}

\begin{document}
	\title{On the Notion of Safe Sliding Mode Control}
	\author{Marco A. Gomez, Christopher D. Cruz-Ancona and Leonid Fridman
		\thanks{M. A. Gomez is with  Department of Mechanical Engineering, DICIS Universidad de Guanajuato, Salamanca, Gto, México (e-mail: marco.gomez@ugto.mx).}
		\thanks{C. D. Cruz-Ancona is with Intelligent Systems Research Lab, Intel Tecnolog\'ia de M\'exico, Intel Labs Zapopan, Jal., Mexico (e-mail: christopher.cruz.ancona@intel.com)}
		\thanks{L. Fridman is with the Department of Robotics and Control, Engineering Faculty, Universidad Nacional Aut\'onoma de M\'exico (UNAM), Mexico 04510, Mexico (e-mail: lfridman@unam.mx)}}

 \maketitle
\pagenumbering{arabic}
	\begin{abstract}
		Within the framework of Sliding Mode Control,  safety critical control {looks at} two problems of theoretical  significance:  construct a sliding manifold that does not intersect   {a given  unsafe} set of the state space, and design a robust controller that safely {takes} the system trajectory to the manifold during the reaching phase {and enforces the sliding motion}. In this technical note, we address both problems. We introduce  the notion of Safe Sliding Manifold (SSM) of relative degree one, and show that it is possible to construct it from the gradient of a class of Lyapunov-like energy functions, previously reported in the literature, used to asses stability and safety of a nominal system. The constructed SSM allows us providing a robust safe controller for a fair general class of uncertain nonlinear systems.        
	
	\end{abstract}
	\begin{IEEEkeywords}
		Safe sliding mode control; Robust stabilization; Lyapunov redesign; Control Lyapunov functions.   
	\end{IEEEkeywords}
	\section{Introduction}

	Safety critical control, which looks for the design of controllers to satisfy constraints in the system state space,  has been subject of study since the 80s; see e.g. \cite{Leitmann1983,hokayem2007cooperative,hokayem2010coordination, blanchini1999set,aubin}. Application of this class of controllers is of particular interest, e.g., in adaptive cruise control \cite{Amesetal2016} and obstacle evasion problems in robotics \cite{Molnar2021model}.  In this regard, taking into account potential disturbances and uncertainties   becomes critical  as they might jeopardize the safety and stability properties of the system.
	
\begin{figure}[h]
		\centering
		\includegraphics[scale=0.35]{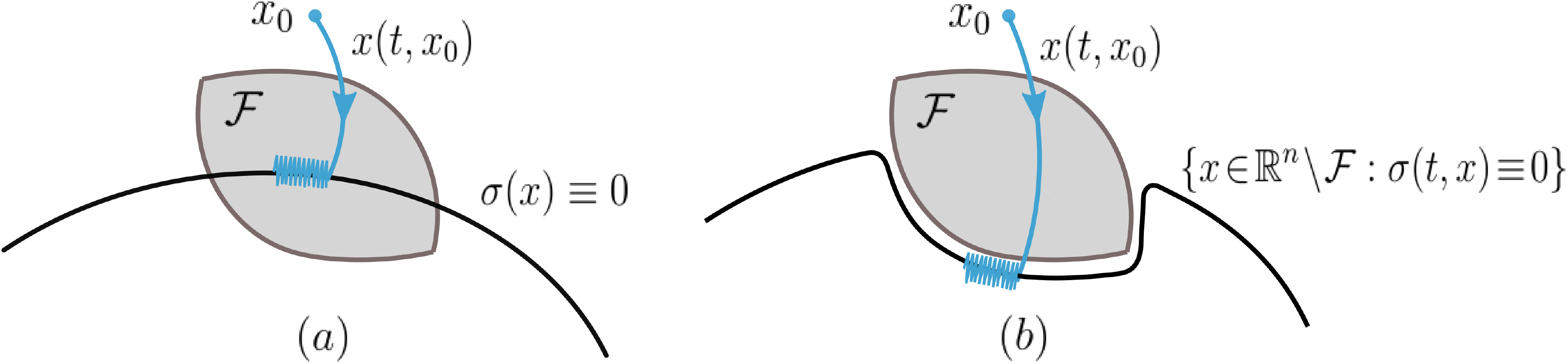}
		\caption{An unsafe trajectory (in blue) converging to: (a) a classical sliding manifold piercing an unsafe set  and (b) a sliding manifold taking into account restrictions in the state space to avoid the unsafe set.}
		\label{fig:sssd}
	\end{figure}

	Sliding Mode Control (SMC), which provides theoretical exact compensation of uncertainties and disturbances during the sliding mode, has been a powerful approach used  to design robust controllers since the 70s. Different methodologies  for the design of sliding manifolds are known in the literature, including, for instance, eigenvalue placement and quadratic minimization based approaches \cite{utkin2013sliding, ackermann1998sliding,edwards1998sliding}. None of the reported design methodologies up to now, however, takes into consideration a sliding manifold design with prescribed constraints in the state space. Under the standard scheme of SMC, there is no guarantee that the robust design provides safety of the system. If the proposed sliding manifold  intersects the set of unsafe states, because of its attractiveness property, the trajectories of the system would  necessarily pass through the prohibitive set {while enforcing a sliding mode}. Furthermore, even if the sliding manifold has no intersection with the unsafe set, there is no guarantee that  the trajectories avoid it during the reaching phase;  Figure \ref{fig:sssd} illustrates these cases.
	
		The above gives rise to two issues of theoretical significance not addressed before in the literature: \emph{the construction of  a sliding manifold that  does not intersect a prescribed set of the state space}, and \emph{the characterization of robust controllers that take the trajectories of the system safely to the constructed sliding manifold and keep them therein for all future time}. We address both. To overcome the issue on the intersection of the sliding manifold with the unsafe set, we introduce the notion of Safe Sliding Manifold (SSM) of relative degree one, which is a stable manifold defined on the safe partition of the state space, that is
	\begin{equation*}
		\mathcal{S}_{\textup{sf}}:=\left\{x\in \mathbb{R}^n\setminus \mathcal{F}: \sigma(t,x)\equiv 0\right\},
	\end{equation*}
	where $\sigma$ is a continuous function of the state vector $x$ and $\mathcal{F}$ is an open set containing  the unsafe region but excluding the origin of the system. 	Inspired by Lyapunov redesign ideas \cite{Leitmann1979,Khalil1992,Corless1987adaptive}, we show that a {stable} SSM  of relative degree one can be constructed from the gradient of a class of Lyapunov-like energy functions used in safety control design of nominal systems \cite{RomdlonyJayawardhana2016}. 
	
	While properties of stability and safety of the nominal design are recovered on the SSM by its definition, safety of the trajectories before reaching the SSM is accomplished by incorporating a transient time function to the sliding variable.  {Analogous to finding a suitable path avoiding obstacles in motion planning algorithms, the}  transient time function must be constructed such that it avoids the unsafe set,  it equals to the sliding variable in zero and it is zero after a finite time. The incorporation of the transient time function to the sliding variable enables the proposal of a unit control able to safely take the trajectories from its initial condition to the SSM  in finite time and keep them therein in spite of uncertainties and disturbances.

	The construction of the SSM is  grounded on two assumptions. The first one is that there exists a nominal control design that, for a given initial condition, makes the system solution asymptotically stable and safe (see Definition \ref{def:safe} in the next section),  and that a Lyapunov-like energy function associated to the closed-loop nominal system is available. Despite this assumption might seem strong, it is possible to construct a nominal control and an associated  energy function following the methodology proposed in \cite{RomdlonyJayawardhana2016}, at least for a subset of admissible initial conditions \cite{BraunKellet2020comment}, relying on the combination of a Control Lyapunov Function (CLF) and a Control Barrier Function (CBF)\cite{WielandaAllgower2007}.  The second assumption is that the manifold is connected. Fulfillment of the latter depends on the class of proposed control barrier and Lyapunov  functions, and can be ruled out in some cases. We show, with systems of practical interest, that these assumptions are satisfied.

	The robust safety problem   has also been addressed  by controllers that ensure an Input to State Safety (ISSf) property of the system in \cite{KolathayaAmes2018}, \cite{RomdlonyJayawardhana2019robustness}. While ISSf controllers guarantee that the system trajectories remain within an invariant compact set in spite of the disturbances, they induce a lost of the stability properties of the nominal design. Under other approaches, mostly dedicated to reducing the computational burden for the control synthesis, the problem is targeted via the solution of optimization and optimal control problems; see e.g. reach-avoidance controllers proposed in \cite{fan2021controller} and references therein. 
	
	Recent results introduced in \cite{ren2022razumikhin} keeps similarities with the presented proposal, which was  independently abstracted.  There, SMC theory is used  for guaranteeing  safety and stability of time-delay systems, but no robustness properties of the proposed control scheme are studied.	
	
	A preliminary version of this paper was presented with sketched proofs in \cite{gomez2022cce}, where, as a controller design solution during reaching phase,   it is proposed a variable structure controller (allowing multiple switchings among structures) to avoid reaching a vicinity of the unsafe set. In this paper we present a much more simpler controller design during reaching phase by incorporating {the} previously described transient function. 
	
	The paper is organized as follows. The  preliminaries are introduced in Section \ref{sec:probf}. The construction of the sliding manifold and a  methodology ensuring safe SMC is presented in Section \ref{sec:ssmc}. The theoretical findings are illustrated with numerical examples in Section \ref{sec:example}. Finally, discussion and conclusions are given in Section \ref{sec:conclusion}.\\

	\noindent \textbf{Notation.} For a given set $A\subset \mathbb{R}^n$, $[A]$ denotes the closure and $\partial A$ represents the boundary of the set. The gradient of a function  $f:\mathbb{R}^n\rightarrow \mathbb{R}$ is denoted by $\nabla f(x)\in \mathbb{R}^{1\times n}$ and the Jacobian matrix of a function $h:\mathbb{R}^n\rightarrow \mathbb{R}^m$ by {$\dfrac{\partial h}{\partial x}={\tiny\begin{bmatrix}\nabla h_1(x)\\  \vdots \\ \nabla h_m(x) \end{bmatrix}}\in \mathbb{R}^{m\times n}$, where $h_i(x)$ is the $i$-th component of $h(x)\in \mathbb{R}^m$.} The identity matrix, whose dimensions are clear from the context, is simply denoted by $I$. We use $\|\cdot\|$ for representing the Euclidian norm of both vectors and matrices and $\lambda_{\min}(C)$ for denoting the minimum eigenvalue of a matrix $C$.

	\section{Preliminaries}\label{sec:probf}
	
	We consider systems of the form
	\begin{equation}
		\label{ec:sys}
		\begin{split}
			\dot x(t)=&f(x)+g(x)((I+\Delta_g(t,x))u(t)+\delta(t,x)),\\
			x(0)=&x_0,
		\end{split}
	\end{equation}
	where $x(t)\in \mathbb{R}^n$ and $u(t)\in \mathbb{R}^m$ are the state and the control input, respectively,  $f:\mathbb{R}^n\rightarrow \mathbb{R}^n$ and $g:\mathbb{R}^n\rightarrow \mathbb{R}^{n\times m}$ are locally Lipschitz, {$f(0)=0$},  $\Delta_g:[0,\infty)\times \mathbb{R}^n\rightarrow \mathbb{R}^{m\times m}$ and $\delta:[0,\infty)\times \mathbb{R}^n\rightarrow \mathbb{R}^m$, which characterize the uncertain terms and disturbances, are measurable functions  in $t$ and continuous in $x$. Throughout this paper, we suppose that the uncertain and disturbance terms are bounded, i.e. the functions $\delta$ and $\Delta_g$ satisfy 
	
	\begin{enumerate}
		\item $\|\delta(t,x)\|\leq \rho(t)$ for all $(t,x)\in [0,\infty)\times \mathbb{R}^n$, where  $\rho:[0,\infty) \rightarrow \mathbb{R}$  is a known measurable function.
		
		\item $\|\Delta_g(t,x)\|\leq \varepsilon<1$ for all $(t,x)\in [0,\infty)\times \mathbb{R}^n$ for some positive real number $\varepsilon$. Moreover, for $\mu>-1$,
		\begin{equation*}
			\lambda_{\min}\left(\dfrac{1}{2}(\Delta_g(t,x)+\Delta_g^T(t,x))\right)\geq\mu\: \forall (t,x)\in [0,\infty)\times \mathbb{R}^n.
		\end{equation*}
	\end{enumerate}
	We denote by $x(t,x_0) $ the solution of system \eqref{ec:sys} at time $t$ with $x(0)=x_0$. If there is no confusion from the context, then we simply denote it by $x(t)$.

	We aim at designing a sliding manifold that has no intersection with a given unsafe set in the state space and that enables the design of a robust control scheme that guarantees safe and stability of the uncertain system. Let us consider the nominal system 
	\begin{equation}
		\label{ec:sys_nom}
		\begin{split}
			\dot x(t)=&f(x)+g(x)u_{nom}(t)\\
			x(0)=&x_0.
		\end{split}
	\end{equation}

	\begin{defi}\cite{RomdlonyJayawardhana2016}
		\label{def:safe}
		Let $\mathcal{X}_0\subset \mathbb{R}^n$ {be the set} of initial conditions containing the origin. For a given $x_0\in \mathcal{X}_0$, a solution $x(t,x_0)$ of \eqref{ec:sys_nom} ({resp. of  \eqref{ec:sys}}) in closed-loop with $u_{nom}$ ({resp. $u$})  is said to be safe if  $x(t,x_0)\notin [\mathcal{D}]$ for all $t\geq 0$, where the set $\mathcal{D}$, so-called unsafe set, is open and  $\mathcal{D}\cap \mathcal{X}_0=\emptyset$. 	If $x(t,x_0)$ is safe for any $x_0\in \mathcal{X}_0\subset \mathbb{R}^n$ then the corresponding system is said to be safe. 
	\end{defi}

	{The problem of rendering closed-loop system \eqref{ec:sys_nom}  asymptotically stable, i.e. $x=0$ is Lyapunov stable and $\lim_{t\rightarrow \infty} \|x(t)\|=0$, and safe has been approached by different routes; see, e.g. \cite{RomdlonyJayawardhana2016,OngCortes2019universal} and references therein. We rely on the results reported in \cite{RomdlonyJayawardhana2016}. The fundamental assumption from which we depart is the following:}
	\begin{assum}
		\label{ass:main_ass}
		Let $\mathcal{D}_0$ be a bounded open set containing the unsafe set $\mathcal{D}$ and $x_0\in\mathbb{R}^n\setminus \mathcal{D}_0$ be a given initial condition. There exists a radially unbounded, lower bounded and two-times continuously differentiable function $W_0:\mathbb{R}^n\rightarrow \mathbb{R} $  such that 
		\begin{enumerate}
			\item $W_0(x)>0$ for all $x\in \mathcal{D}_0$.
			\item $\mathcal{U}:=\{ x\in \mathbb{R}^n: W_0(x)\leq 0\}\neq \emptyset$.
			\item $[\mathbb{R}^n\setminus (\mathcal{D}_0 \cup \mathcal{U} )]\cap [{\mathcal{D}_0}]=\emptyset.$
			\item  There exists a nominal control $u_{nom}$ satisfying $\nabla W_0(x)\cdot(f(x)+g(x)u_{nom}(t))< 0$ $\forall x(t,x_0)\in \mathbb{R}^n\setminus (\mathcal{D}_0\cup \{0\})$.
			%		\item \textcolor{red}{For $x(t)=x(t,x_0)$, $x_0\in \mathbb{R}^n\setminus \mathcal{D}_0$, $\nabla W_0(x)\cdot(f(x)+g(x)u_{nom}(t))< 0$ if and only if $x(t)\equiv 0$.}
		\end{enumerate}
	\end{assum}
	
	Assumption \ref{ass:main_ass}, motivated from Proposition 1 in \cite{RomdlonyJayawardhana2016}, ensures that the trajectory $x(t,x_0)$ of closed-loop system \eqref{ec:sys_nom}  is safe and asymptotically stable. In contrast with \cite{RomdlonyJayawardhana2016}, it requires functions that are two-times continuously differentiable, which shall enable us to define a continuous function as a sliding variable in the forthcoming section.

	{Proposition 3 in \cite{RomdlonyJayawardhana2016} provides a constructive formula for $W_0$ relying on the combination of a CLF and a CBF, from which a continuous control law {for the nominal system  akin to} Sontag's formula is obtained. However, as shown in \cite{BraunKellet2020comment},  the proposed nominal control {ensures} that only some system trajectories, not all,  are safe and stable. In view of this and since we strongly rely on the class of constructed functions in \cite{RomdlonyJayawardhana2016},  we restrict ourselves { in item 4 of the assumption to a specific trajectory for a given initial condition}.  The nominal design for achieving safety and stability properties of the system (i.e. for any $x_0\in \mathcal{X}_0$) is still under active  research and is beyond the proposal of this note.}

	\section{Safe Sliding Mode Control}\label{sec:ssmc}
	\label{sec:sf_smc}	Grounded on Assumption 1, we show in this section that a safe sliding manifold can be constructed from the  gradient of function $W_0$.  We take inspiration from the Lyapunov-Redesign idea: For any given $x_0$ for which Assumption 1 holds, we have that the time derivative of the function $W_0$ along the solution $x(t, x_0)$ of system (1) with $u(t)=u_{nom}(t)$ satisfies 
	\begin{multline}
		\label{ec:time_derivative}
%	%	\begin{split}
			\left.\dfrac{d}{dt}W_0(x(t))\right|_{\eqref{ec:sys}}=\nabla W_0(x)\cdot(f(x)+g(x)u_{nom}(t))+\nabla W_0(x)\cdot g(x)\left(\Delta_g(t,x)u_{nom}(t)+\delta(t,x)\right)\\	< \nabla W_0(x)\cdot g(x)\left(\Delta_g(t,x)u_{nom}(t)+\delta(t,x)\right)
%	%	\end{split}
	\end{multline}
$\forall x(t,x_0)\in \mathbb{R}^{n}\setminus (\mathcal{D}_0\cup \{0\}).$ The second term in the right hand side that contains the uncertainties, referred as the residual term, induces an indefinite sign of the time-derivative, preventing to assess whether the system trajectory is stable and safe or not. Notice however that the obtained time-derivative already suggests a control scheme to recover the negative sign. Namely, adding a robustifying control to the nominal one to  force the sign of the residual term to be negative or to be zero, i.e. considering a control law of the form
\begin{equation}
\label{eq:nomprob}
    u(t)=u_{nom}(t)+u_{rob}(t)
\end{equation}
to have 
\begin{equation*}
\left.\dfrac{d}{dt}W_0(x(t))\right|_{\eqref{ec:sys}}<\nabla W_0(x)\cdot g(x)\left(\Delta_g(t,x)u(t)+u_{rob}(t)+\delta(t,x)\right) 
\end{equation*}
{Designing $u_{rob}(t)$ to force the residual   term to be negative is the core idea of the Lyapunov-Redesign approach, to force it to be zero is our proposal, which leads to the construction of a sliding manifold that naturally adopts the representation $\nabla W_0(x)\cdot g(x).$} 

The section is divided into two parts. In the first one we discuss the construction of a SSM departing from the previously mentioned {representation}, and in the second one we exploit it in order to propose a robust controller for guaranteeing a safe and stable solution of the system. 
	\subsection{A safe sliding manifold}

\begin{figure}
		\centering
		\includegraphics[scale=0.24]{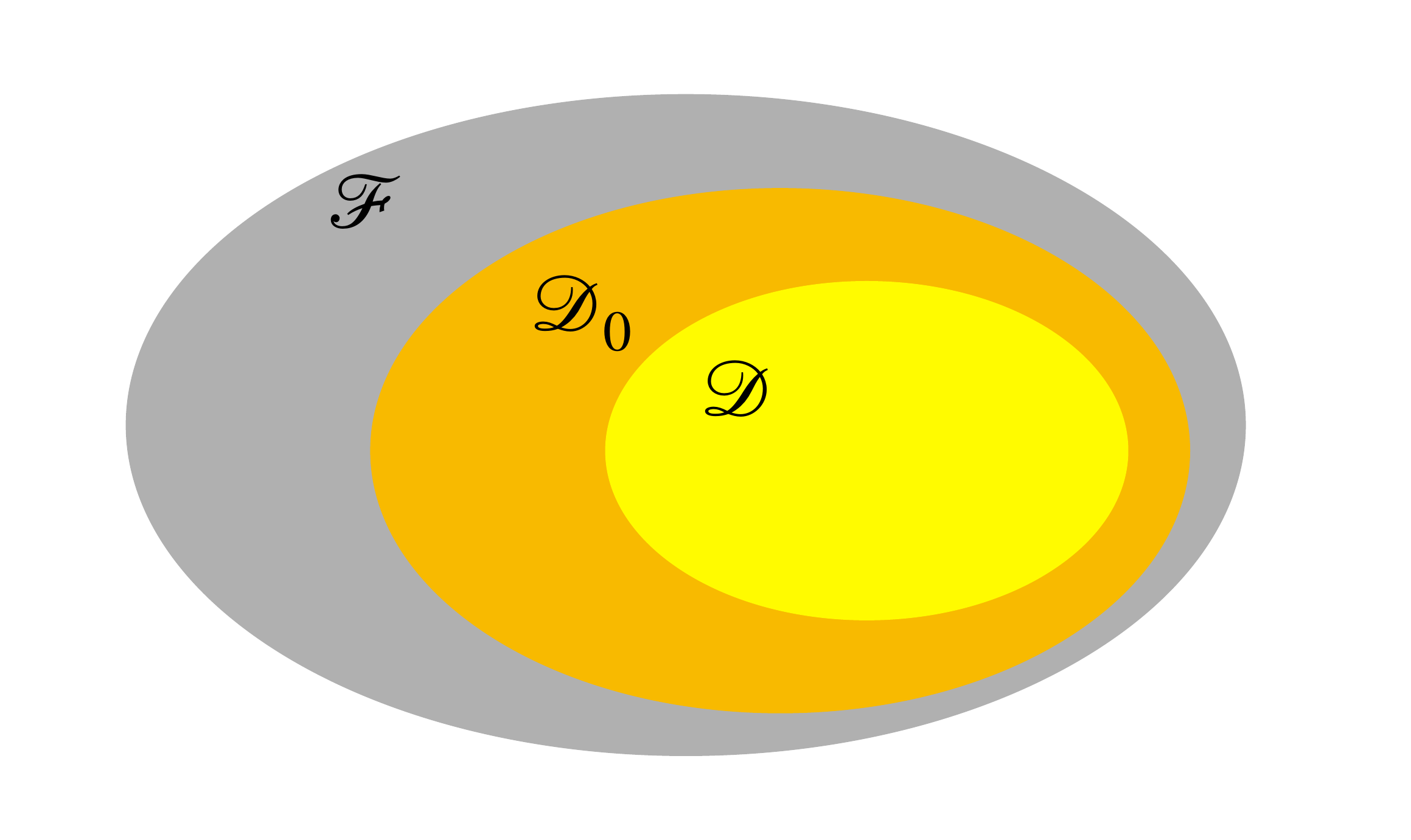}
		\caption{Set inclusions  $\mathcal{F}$, $\mathcal{D}_0$ and $\mathcal{D}$.}
		\label{fig:sets}
	\end{figure}

	The proposed sliding variable to construct a SSM, $\mathcal{S}_{\textup{sf}}$, is given by
	\begin{equation*}
		\sigma(t,x):=\nabla W_0(x)\cdot g(x), \:x\in \mathbb{R}^n\setminus \mathcal{F},
	\end{equation*}
	which is continuous by Assumption \ref{ass:main_ass} and $\mathcal{F}$ is a constructed set 
	such that $\mathcal{D}\subset \mathcal{D}_0\subseteq \mathcal{F}$ and that it does not contain the origin; see Figure \ref{fig:sets}. Enlarging the unsafe set is standard to make the set avoidable, cf. with \cite{Corless1987adaptive}.  Hereafter, if no confusion arises, we just consider the time argument for the function $\sigma$, i.e. $\sigma(t)=\sigma(t,x)$.

	 Since $\mathcal{D}\subset \mathcal{F}$, it holds by definition that  $\mathcal{S}_{\textup{sf}}\cap \mathcal{D}=\emptyset$.  Let us remark next some subtleties that justifies the necessity of constructing the set $\mathcal{F}$. Owing to the nature of the energy function $W_0$, for $x\in \mathbb{R}^n$, there is no guarantee that  the manifold $$\left\{\nabla W_0(x) \cdot g(x)=0\right\}$$
	avoids the unsafe set $\mathcal{D}$ nor that  is connected. Consider the following example taken from \cite{RomdlonyJayawardhana2016}. 
	\begin{example}
		\label{ex:Jaya_ex}
		Let  $\mathcal{D}$ be the unsafe set of a two-dimensional system be characterized by \begin{equation}
			\label{ec:unsafe_set_ex}
			\mathcal{D}=\left\{x\in \mathcal{X}:h(x)<4 \right\}.
		\end{equation}
		The system corresponds to the one in Example 1 in \cite{RomdlonyJayawardhana2016}, revisited in Section \ref{sec:example} in this note. A CLF and a CBF for this system  are
	$V(x)=x_1^2+x_1x_2+x_2^2$ and 
		\begin{equation*}
			\mathcal{B}(x)=\left\{\begin{array}{cc}
				e^{-h(x)}-e^{-4}, & \forall x\in \mathcal{X}\\
				-e^{-4}, & \text{otherwise}
			\end{array}\right.,
		\end{equation*}
		respectively, where $h(x)=\dfrac{1}{1-(x_1-2)^2}+\dfrac{1}{1-x_2^2}$ and  $\mathcal{X}=(1,3)\times (-1,1)$ . 
		The function $W_0(x)$ is then constructed as
		\begin{equation}
			\label{ec:function_W_ex}
			W_0(x)=V(x)+\lambda \mathcal{B}(x)+\kappa,
		\end{equation}
		with $\lambda=1000$ and $\kappa=-0.7$. Direct calculations lead to the directional derivative
		\begin{equation*}
			\nabla W_0(x)\cdot e_2=\left\{\begin{array}{cc}
				x_1+2x_2-2\lambda x_2\dfrac{e^{-h(x)}}{(x_2^2-1)^2} &  \forall x\in \mathcal{X}\\
				x_1+2x_2, & \forall x\in \mathbb{R}^n\setminus \mathcal{X}
			\end{array}\right.,
		\end{equation*}
		with $e_2=\begin{pmatrix}
			0 & 1
		\end{pmatrix}^T$.
		In Figure \ref{fig:unsafe_set} (left panel), where the set $\mathcal{D}$ is fill in yellow and the level curve $\nabla W_0(x)\cdot e_2=0$ is depicted by the black continuous line, it is clearly observed that the set $\left\{x\in \mathbb{R}^n:\nabla W_0(x)\cdot e_2=0 \right\}$ is  disconnected and there is a subset of it that belongs to the unsafe set $\mathcal{D}$. 
		\begin{figure}
			\centering
			\includegraphics[scale=1.2]{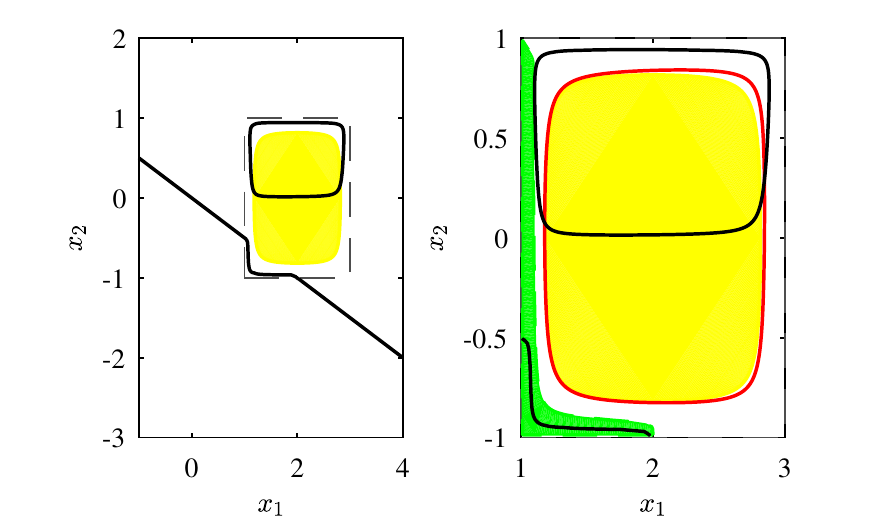}
			\caption{Left panel: $\partial \mathcal{X}$ (dash line in black), set $\mathcal{D}$ (fill in yellow) and manifold $\{\nabla W_0(x)\cdot e_2=0\}$ (continuous line in black) in the plane $(x_1,x_2)$. Right panel: Zoom of the figure on the left, with $\mathcal{X}\setminus \mathcal{F}$ fill in green and $\partial \mathcal{D}_0$ depicted in red, where $\mathcal{D}_0=\{x\in \mathcal{X}: W(x)>0\}$. The details of the computation of $\mathcal{F}$ are presented in Section \ref{sec:example}.}
			\label{fig:unsafe_set}
		\end{figure}
	\end{example}
	
	The above motivates the following assumption: 
	\begin{assum}
		\label{ass:connected}
		The set $\mathcal{S}_{\textup{sf}}$ is connected and contains the origin.
	\end{assum}
	
	That the sliding manifold satisfies the above assumption depends on the function $W_0$ and a suitable choice of the set $\mathcal{F}$. For the class of proposed energy functions in \cite{RomdlonyJayawardhana2016}, recalled here at the beginning of Section \ref{sec:example}, satisfying Assumption \ref{ass:main_ass},  it is possible to ensure that the origin is contained in the manifold as $\nabla W_0(x)=0$ for $x=0$, and that what might induce a disconnected sliding manifold clearly is the CBF defined on the subset $\mathcal{X}$.
	
	If the manifold
	$\{\nabla W_0(x)\cdot g(x)=0\}$ intersects the unsafe set, as in the above example, a careful construction of the set $\mathcal{F}$ is required. The complexity of such construction relies on the fact that  $\mathcal{F}$ must  contain the subset of the manifold that is unsafe while excluding the whole subset of the  manifold containing the origin. In the right panel of Figure \ref{fig:unsafe_set}, the white region corresponds to the set $\mathcal{F}$ and the green region to the set $\mathcal{X}\setminus \mathcal{F}$, on  $(1,3)\times (-1,1)$. The figure shows that the set  $\mathcal{F}$ barely excludes the stable and safe  sliding manifold. The complexity of the construction of $\mathcal{F}$, however, might be significantly reduced {with a proper reaching phase design};  see Remark \ref{rem:drop_ass} in the next subsection.

	If $\{\nabla W_0(x)\cdot g(x)=0\}\cap \mathcal{D}_0=\emptyset$, on the other hand, the construction of $\mathcal{F}$ is straightforward as the manifold is already safe. In this case, if $W_0$ is constructed as proposed in \cite{RomdlonyJayawardhana2016}, i.e the combination of a CLF and a CBF, then we can set  $\mathcal{F}=\mathcal{X}$, where $\mathcal{X}$ is a compact and connected set on which the CBF is defined, and the construction of the SSM is deduced only from  the gradient of the CLF. This is illustrated by Example 3. The above gives way to the following remark:
	
	\begin{rem}
	\label{rem:CLF}
	    If there is a CLF $V(x)$ for the nominal system  such that $\{\nabla V(x)\cdot g(x)=0\}\cap \mathcal{D}=\emptyset$, then Assumption \ref{ass:main_ass} and Assumption \ref{ass:connected} can be ruled out as the stable SSM can be constructed  from the CLF. It is straightforward to prove that the complete robust scheme developed in the coming subsection still provides safe and stable solutions, cf. Lemma \ref{lem:conver_S} and Theorem \ref{theo:stability_safe}. 
	\end{rem}
	
	It is worthy of mention that, even in the stabilization problem context, up to the best of the authors' knowledge, there are no reported results in the literature that exploit the structure of CLF to design stable sliding manifolds.

	\subsection{Robust stabilization and safety}
	\label{sec:robusttotal}
	
	From now on we assume that the matrix $g_0(x):=\dfrac{\partial \sigma}{\partial x}g(x)$, $g_0:\mathbb{R}^n\rightarrow \mathbb{R}^{m\times m}$, is non-singular for all $x\in \mathbb{R}^n$, i.e.  the relative degree of the sliding variable dynamics is well-defined and equal to one. It is clear that $\sigma(t)\equiv 0$ induces a negative sign in the time-derivative \eqref{ec:time_derivative}. We rely on the ideas of SMC in order to achieve that in two stages. The first, so called reaching phase, consists in taking the trajectory to the sliding manifold, whereas the second corresponds to the evolution of the solution on it. 
	
	While the dynamics on the sliding manifold are already safe since by definition $\mathcal{S}_{\textup{sf}}\cap \mathcal{D}=\emptyset$, nothing can we say on safety of the trajectories in the reaching phase, in which the trajectories might pass through the unsafe set. The robust controller aiming at guaranteeing that a solution of the closed-loop system is safe in the presence of uncertainties and disturbances is given by
	\begin{equation}
		\label{ec:robust_control}
		u_{rob}(t)=-k(t,x)\dfrac{{\bar \sigma_{sf}^T(t)}}{\|\bar \sigma_{sf}(t)\|},
	\end{equation}
	where
	\begin{equation*}
		\bar \sigma_{sf}(t):=\left\{\begin{array}{cc}
			\bar \sigma(t)-{\phi(t)g_0(x)}, & 0\leq t< t_f\\
			\bar \sigma(t), & t\geq t_f
		\end{array}\right.,
	\end{equation*}
	$t_f>0$ is related with the initial conditions as discussed below, {$\bar \sigma(t) =\sigma(t)g_0(x)\in \mathbb{R}^{1\times m}$} and the gain $k(t,x)$  is selected such that
	\begin{equation*}
		k(t,x)\geq \dfrac{\rho_0(t,x)+\|g_0^{-1}(x)\dot \phi^T(t)\|+q}{1+\mu},
	\end{equation*}
	with $q>0$ and
	\begin{equation*}
		\rho_0(t,x)=\left\| g_0^{-1}(x)\dfrac{\partial \sigma}{\partial x}\left(f(x)+g(x)u_{nom}(t)\right)\right\|+\varepsilon \|u_{nom}(t)\|+\rho(t).
	\end{equation*}

The function {$\phi:[0,\: \infty) \rightarrow \mathbb{R}^{1\times m}$}, constructed for a given initial condition, is  continuous and satisfies the following properties:
	\begin{enumerate}[P.1.]
		\item Its image is the set $\Sigma=\{d\in \mathbb{R}^{1\times m}:\sigma(t,x)=d,\:\:x\in\mathbb{R}^n\setminus \mathcal{F}_r\}$, where $\mathcal{F}_r\supseteq \mathcal{F}$,
		\item $\phi(0)=\sigma(0)$ and 
		\item $\phi(t)=0$ for all $t\geq t_f$.
	\end{enumerate}
	The idea of the robust controller proposal, which arises  from the integral sliding mode approach, see e.g. \cite{levant2007integral,utkin1996integral,matthews1988decentralized},  consists in ensuring that, on $t\in[0,t_f)$, $\sigma(t)$ follows the planned trajectory {given} by $\phi(t)$, i.e. $\sigma(t)=\phi(t)$  $\forall \:t\in[0,t_f)$, and that $\sigma(t)=0$ $\forall \:t\geq t_f$. The property P.1 of $\phi(t)$ intends to guarantee safety of the reaching phase when $\phi$ is followed by $\sigma$. 
	
	{Since the robustifying term in \eqref{ec:robust_control} is discontinuous, any solution of the closed-loop system \eqref{ec:sys}-\eqref{eq:nomprob} is  considered in the sense of Filippov \cite{filippov1988}.} 
	\begin{lem}
		\label{lem:conver_S}
		For a given $x_0\in \mathbb{R}^n\setminus \mathcal{D}_0$, suppose that Assumption \ref{ass:main_ass}  and  Assumption \ref{ass:connected}  are satisfied, and that  there is a function $\phi$ satisfying Properties P.1, P.2 and P.3.  The solution $x(t,x_0)$ of system \eqref{ec:sys} in closed-loop with $u(t)=u_{nom}(t)+u_{rob}(t)$   converges to the sliding manifold $\mathcal{S}_{\textup{sf}}$ in a finite time $t_f$ and $x(t,x_0)\notin \mathcal{D}$ for all $t\in [0,t_f]$.
	\end{lem}
	
	\begin{proof}
		It is enough to prove that $\sigma(t)-\phi(t)=0$ for all $t\geq 0$. In fact, if this is the case, it follows that $\sigma(t)=\phi(t)$ for $t\in [0,t_f)$ and $\sigma(t)=0$ for $t\geq t_f$. Moreover, by definition of the image of $\phi$,  $x(t,x_0)\notin \mathcal{D}$ for all $t\in [0,t_f]$. 
		
		The derivative of the sliding variable along the solutions of system \eqref{ec:sys} is given by
		\begin{equation*}
			\dot \sigma (t)=\dfrac{\partial \sigma}{\partial x}(f(x)+g(x)u_{nom}(t))+g_0(x)\left((I+\Delta_g(t,x)u_{rob}(t)+\Delta_g(t,x)u_{nom}+\delta(t,x))\right),
		\end{equation*}
		for $x\in \mathbb{R}^n\setminus \mathcal{F}$ and the closed-loop dynamics  are written as
		\begin{equation}
			\label{ec:sigma_sys}
			\begin{split}
				\dot\sigma (t)=&g_0(x)\left((I+\Delta_g(t,x))u_{rob}(t)+\delta_0(t,x)\right),\quad x\in \mathbb{R}^n\setminus \mathcal{F}\\
				\sigma(0)=&\sigma_0,
			\end{split}
		\end{equation}
		where
		\begin{equation}
		\label{ec:delta0}
			\delta_0(t,x)=g_0^{-1}(x)\dfrac{\partial \sigma}{\partial x}\left(f(x)+g(x)u_{nom}(t)\right)+\Delta_g(t,x)u_{nom}(t)+\delta(t,x).
		\end{equation}
		It follows from the assumed  bounds of the disturbance terms that
		\begin{equation}
			\|\delta_0(t,x)\|\leq \rho_0(t),\quad \forall t\in [0,\infty).
		\end{equation}
		Let us consider the Lyapunov function
		\begin{equation*}
			V_{smc}(t)=\dfrac{1}{2}\|\sigma(t)-\phi(t)\|^2.
		\end{equation*}
		The time-derivative of the function along solutions of system \eqref{ec:sigma_sys} satisfies
		\begin{multline*}
%			\begin{split}
				\dfrac{d}{dt}V_{smc}(t)=-k(t,x) {\Vert \bar \sigma_{sf}(t)\Vert}  -\dfrac{k(t,x)}{\|\bar \sigma_{sf}(t)\|}{\bar \sigma_{sf}(t)}\left(\dfrac{1}{2}(\Delta_g(t,x)+\Delta^T_g(t,x))\right) {\bar \sigma_{sf}^T(t)} +{\bar \sigma_{sf}(t)}\delta_0(t,x)-{\bar\sigma_{sf}(t)}g_0^{-1}(x)\dot \phi^T(t)\\
				\leq -k(t,x)\|\bar \sigma_{sf}(t)\| \left(1+\lambda_{\min}\left(\dfrac{1}{2}(\Delta_g(t,x)+\Delta^T_g(t,x))\right)\right)+\|\bar \sigma_{sf}(t)\|\|\delta_0(t,x)\|+\|\bar \sigma_{sf}(t)\|\|g_0^{-1}(x)\dot \phi^T(t)\|\\
				\leq -k(t,x)(1+\mu)\|\bar \sigma_{sf}(t)\|+\|\bar \sigma_{sf}(t)\|\left( {\rho_0(t)}+\|g_0^{-1}(x)\dot \phi^T(t)\|\right) \leq -q \|\bar \sigma_{sf}(t)\|,\quad \forall(t,x)\in [0,\infty)\times  \mathbb{R}^n\setminus \mathcal{F}_r.
%			\end{split}
		\end{multline*}
		The definition of the domains is consistent since $\mathcal{F}\subseteq\mathcal{F}_r$. As $\bar \sigma(t) -g_0(x)\phi(t)=(\sigma(t)-\phi(t))g_0(x)$,
		\begin{equation*}
			\|\bar \sigma_{sf}(t) \|\leq \sqrt{\lambda_{\min}(g_0(x)g_0^T(x))}\|\sigma(t)-\phi(t)\|=\sqrt{2\lambda_{\min}(g_0(x)g_0^T(x))}\sqrt{V_{smc}(t)},
		\end{equation*}
		and we arrive at
		\begin{equation*}
			\dfrac{d}{dt}V_{smc}(t)\leq -q\sqrt{2\lambda_{\min}(g_0(x)g_0^T(x))}\sqrt{V_{smc}(t)}
		\end{equation*}
		$\forall x\in \mathbb{R}^n\setminus \mathcal{F}_r.$
		By the comparison lemma, the above inequality implies that $\sigma(t)-\phi(t)=0$ for all $t\geq 0$, which completes the proof. \hfill $\square$
		
	\end{proof}

	\begin{rem}
		\label{rem:initial_cond}
		If there is no exact knowledge of the initial condition of the system, P.2. of $\phi$ might not be satisfied, i.e. $\phi(0)\neq \sigma(0)$. In this case, we cannot ensure that $\sigma(t)=\phi(t)$ for all $t\geq 0$, but only for all $t\geq T$, with 
		\begin{equation}
			T=\dfrac{\|\sigma_{sf}(0)\|}{q\sqrt{2}\sqrt{\lambda_{\min}(g_0(x)g_0^T(x))}}.
		\end{equation}
		Regardless whether $T<t_f$ or $t_f<T$, it is guaranteed that $\sigma(t)=0$ for all $t\geq \max\{t_f,T\}$: if $T<t_f$, then the trajectory hits $\phi$ at time $T$ and becomes zero after time $t_f$. If $t_f<T$, then the trajectory hits $\phi$ when it is already zero. In this case, the problem of keeping the trajectory safe for $t\in [0,\max\{t_f,T\}]$ re-emerge. Under the assumption that there is, at least, knowledge of a sufficiently small ball containing the unknown initial condition and the distance of this ball to the unsafe set is sufficiently large, by setting $\phi(0)$ within this ball, by continuity of the solution the trajectory shall remain out of the unsafe set until it hits the SSM. 
	\end{rem}
	
	\begin{rem}
		\label{rem:drop_ass}
		The requirement of $\mathcal{S}_{\textup{sf}}$  to be connected can be relaxed by constructing $\mathcal{F}_r$,  cf. P.1. of $\phi(t)$, larger than any given $\mathcal{F}$ and such that avoids the origin. This ensures that the system trajectory does not hit the subset of the manifold that does not contain the origin during the reaching phase, but the one that does contain it. The price to be paid is the introduction of conservatism  as  the set of initial conditions is smaller. This is clearly observed in Example \ref{ex:Jaya_ex}. With  $\mathcal{F}=\mathcal{D}_0\subset \mathcal{D}$, the sliding manifold is safe but not connected  anymore (see the red line in Figure \ref{fig:unsafe_set} denoting the boundary of $\mathcal{D}_0$). With this choice, the unsafe set is avoided, but the rest of closed curve of the manifold is not. However,  by choosing $\mathcal{F}_r=\mathcal{X}\supset \mathcal{D}_0$, by Lemma \ref{lem:conver_S}, one ensures convergence of the trajectory  to the stable sliding manifold. Clearly, it is not possible to choose initial conditions within the set $\mathcal{X}$ anymore.
	\end{rem}

	The next theorem immediately follows from  \Cref{lem:conver_S}.
	\begin{theo}
		\label{theo:stability_safe}
		Suppose that the hypothesis of \Cref{lem:conver_S} are satisfied. The solution $x(t,x_0)$ of system \eqref{ec:sys} in closed-loop with  $u(t)=u_{nom}(t)+u_{rob}(t)$ is safe and asymptotically stable.
	\end{theo}
	\begin{proof}
		By \Cref{lem:conver_S}, $\sigma(t)=\phi(t)\: \forall t\in [0,t_f)$ and $\sigma(t)=\dot \sigma(t)=0\: \forall t\geq t_f$, which implies that the solution in closed-loop with the robust controller is safe. It remains to prove  asymptotic stability. Since from expression \eqref{ec:delta0} we have that
		\begin{equation*}
			\Delta_g(t,x)u_{nom}+\delta(t,x)=\delta_0(t,x)-g_0^{-1}(x)\dfrac{\partial \sigma}{\partial x}\left(f(x)+g(x)u_{nom}\right),
		\end{equation*}
		system \eqref{ec:sys} can be written as
		\begin{equation*}
			\dot x(t)=\left(I-g(x)g_0^{-1}(x)\dfrac{\partial \sigma}{\partial x}\right)\left(f(x)+g(x)u_{nom}(t)\right)+g(x)g_0^{-1}(x)\dot \sigma (t).
		\end{equation*}
		Time-derivative of the function $W_0$ along the solution $x(t,x_0)$ in closed-loop with $u(t)=u_{nom}(t)+u_{rob}(t)$ gives
		\begin{equation*}
			\begin{split}
				\dfrac{d}{dt}W_0(x(t)) &=\nabla W_0(x)\cdot(f(x)+g(x)u_{nom}(t))
				-{\sigma(t)}g_0^{-1}(x)\dfrac{\partial \sigma}{\partial x} (f(x)+g(x)u_{nom}(t))\\
				&=\nabla W_0(x)\cdot(f(x)+g(x)u_{nom}(t))<0,\:\: t\geq t_f,
			\end{split}
		\end{equation*}
		where the last inequality followed from Statement 4 of Assumption \ref{ass:main_ass}, and implies that 
		\begin{equation*}
			W_0(x(t))< W_0(x(t_f))<\infty \quad \forall t\geq t_f.
		\end{equation*}
		Since $W_0$ is a radially unbounded function and the closed-loop solution of \eqref{ec:sys} is absolutely continuous, we have that the trajectories of $x$ are bounded for all $t\geq 0$, i.e. the set $\{\:x(t): \:t\in [0,\infty) \:\}$ is pre-compact. The same arguments presented in the proof of  \cite[Proposition 1]{RomdlonyJayawardhana2016} allows to conclude asymptotic stability of the closed-loop system solution. \hspace*{\fill} $\square$
	\end{proof}

	\section{Numerical examples}
	\label{sec:example}
	
	We present two examples. The first one corresponds to  Example \ref{ex:Jaya_ex} revisited, and the second is taken from \cite{OngCortes2019universal}. The latter illustrates the discussion in Remark \ref{rem:CLF}. For the construction of $W_0$, we rely on Proposition 3 in \cite{RomdlonyJayawardhana2016}. Specifically,  
	\begin{equation*}
		W_0(x)=V(x)+\lambda \mathcal{B}(x)+\kappa
	\end{equation*}
	where $V$ and $\mathcal{B}$ are CLF and CBF,  respectively. The CBF is such that $\mathcal{B}(x)=-\epsilon$, $\epsilon>0$, for $x\in\mathbb{R}^n\setminus \mathcal{X}$, with $\mathcal{X}$ a compact and connected set containing the unsafe set and excluding the origin. The constant $\lambda$ satisfies $\lambda>\dfrac{c_2c_3-c_1c_4}{\epsilon}$ and $\kappa=-c_1c_4$, where $c_1$ and $c_2$ are such that $c_1\|x\|^2\leq V(x)\leq c_2\|x\|^2$, {$c_3=\max_{x\in \partial \mathcal{X}}\Vert x \Vert^2$ and $c_4=\min_{x\in \partial{\mathcal{D}}}\Vert x \Vert^2$.} The nominal control is constructed as
	\begin{equation}
		\label{ec:nominal_control}
		u_{nom}(t)=\left\{\begin{array}{cc}
			- \dfrac{a+\sqrt{a^2+\gamma \|b\|^4}}{{bb^T}}{b^T}& b\neq 0\\
			0 & \text{otherwise}
		\end{array}\right.
	\end{equation}
	where $a:=\nabla W_0(x)\cdot f(x)$, $b:=\nabla W_0(x)\cdot g(x)$ and $\gamma>0$, which is set as $\gamma=2$.
	Assumption \ref{ass:main_ass} holds for the initial states tested, whereas Assumption \ref{ass:connected} is satisfied by correctly constructing the set $\mathcal{F}$. The construction of $W_0$ shows a remarkable fact about considering CLF and CBF within the framework of safe sliding mode control: The SSM is a stable set, i.e. the reduced dynamics once in sliding mode is asymptotically stable even when the nominal controller \eqref{ec:nominal_control} vanishes at the SSM. Indeed, $\frac{d}{dt}W(x(t))=\nabla W_0(x)\cdot f(x)<0$ with $x\in \mathcal{S}_{sf}$.
	
	The behavior of the system response with the nominal control and  the robust control scheme is contrasted in both examples via numerical simulations, for which we use the Euler method with integration step of $0.0001$. In the corresponding figures, we depict in {red and blue} the results obtained with  $u(t)=u_{nom}(t)$ and $u(t)=u_{nom}(t)+u_{rob}(t)$, respectively.
	
	\begin{example}
		\label{ex:ex_Jaya_rev}
		We take up the example proposed in \cite{RomdlonyJayawardhana2016} with uncertain terms, which is in the form \eqref{ec:sys} with
		\begin{equation*}
			f(x)=\begin{pmatrix}
				x_2\\
				-s(x_2)-x_1
			\end{pmatrix}\quad \text{and}\quad g(x)=\begin{pmatrix}
				0\\1
			\end{pmatrix},
		\end{equation*}
		where $s(x_2)=\left(0.8+0.2e^{-100|x_2| }\right)\tanh(10x_2)+x_2$. For the numerical simulation we consider the function $\delta(t)=-5\sin(10t)$ as the disturbance term and $\Delta_g=0$.  The unsafe region in the $(x_1,x_2)$-plane is characterized by the set $\mathcal{D}$ in \eqref{ec:unsafe_set_ex}; see Figure \ref{fig:unsafe_set}. We consider the function $W_0$ given in equation \eqref{ec:function_W_ex} and with this, 
		the nominal control $u_{nom}$ is constructed as in \eqref{ec:nominal_control}.

		{The sliding variable is given by
			\begin{equation*}
				\sigma(t)=\nabla W_0(x)\cdot g(x)=\left\{\begin{array}{cc}
					x_1+2x_2-2\lambda x_2\dfrac{e^{-h(x)}}{(x_2^2-1)^2} &  \forall x\in \mathcal{X}\setminus \mathcal{F}\\
					x_1+2x_2, & \forall x\in \mathbb{R}^n\setminus \mathcal{X}
				\end{array}\right.,
			\end{equation*}
			with $ \mathcal{F}=\left\{x\in \mathcal{X}: W_0(x)> -\eta \right\}$, where  $\eta$ is computed as follows. Let $x\in \mathbb{R}^n\setminus \mathcal{X}$ be such that $\sigma(t)=0$, then $x_1=-2x_2$ and
			\begin{equation*}
				\left\{x\in \partial \mathcal{X}:\sigma(t)=0\right\}=\left\{x\in \mathbb{R}^2: x=(2,-1),x=\left(1,-\dfrac{1}{2}\right) \right\}
			\end{equation*}
			Evaluating the function $W_0(x)$ on  $(x_1,x_2)=(-2x_2,x_2)=(2,-1)\in \partial \mathcal{X}$, we get $W_0(x)=-16.0156$. By continuity of the sliding variable $\sigma$, we can choose $\eta=16.0156$. The set $\mathcal{X}\setminus \mathcal{F}$ is illustrated in Figure \ref{fig:unsafe_set}, right panel, in green.}
		
		We construct  $\phi(t)=\alpha+\alpha\cos(t)+\beta\sin(t)$, where {$\alpha$ and $\beta$} depend on the initial condition of the system. In particular,   for $x_0=[3.5\:\: 1.5]^T$,  $x_0=[2\:\: 1]^T$ and {$x_0=[2.5\:\: 1.25]^T$}, $(\alpha,\beta)=(3,-1)$, $(\alpha,\beta)=(2,2)$ and $(\alpha,\beta)=(2.5,0.75)$, respectively. For the first case, $t_f=2.5$ and for the second and third one $t_f=\pi$. For these values, $\alpha=\sigma(0)/2$ and $\beta$ was chosen such that the unsafe region is avoided.  In Figure \ref{fig:example1_time} we show the results of numerical simulation for the initial condition $x_0=[3.5\:\:\: 1.5]^T$.    For the robust controller we choose $q=0.5$ and $\rho(t)=5$ for all $t\geq 0$. Figure \ref{fig:example1_time} (b) illustrates the main idea of the proposal, which is that {adding the robustifying} controller makes the function $W_0$  decreasing after a finite time $t_f$.
		
		In Figure \ref{fig:example1_states} we display the trajectories in the plane $(x_1,x_2)$, where the trajectories of the perturbed system in closed loop with the robust and nominal controller are depicted on the left and right, respectively. The robust controller is tested for more initial conditions. The results clearly show that the trajectories hit  the safe manifold $\mathcal{S}_{\textup{sf}}$ and after that they remain therein to converge safely and asymptotically to the origin despite the presence of uncertainties and disturbances. %the disturbance term
		
		\begin{figure}
			\centering
			\includegraphics[scale=1.2]{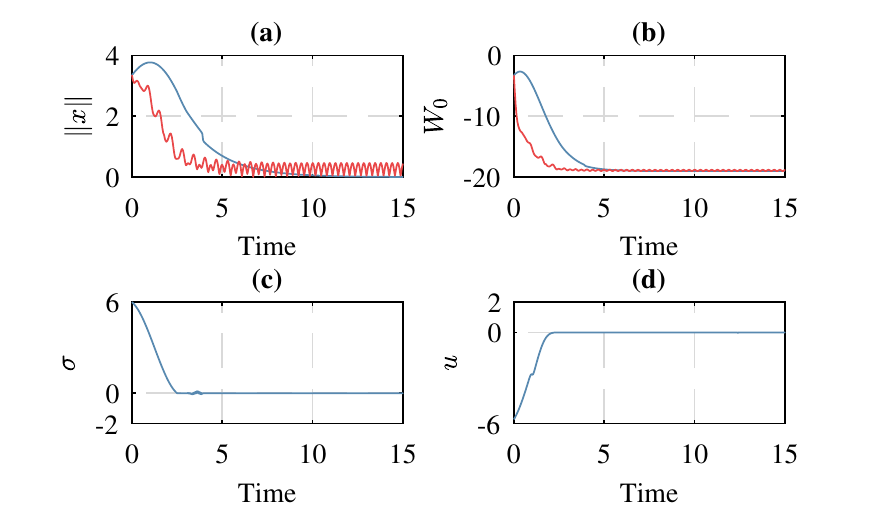}
			\caption{Results of the numerical simulation in Example \ref{ex:ex_Jaya_rev}  with the initial condition $x_0=[3.5 -2.5]^T$. (a) Norm of the solution $x(t,x_0)$; (b) Function $t\mapsto W_0(x(t))$;  (c) Sliding variable $t\mapsto \sigma(t)$; (d) Control signal $u(t)=u_{nom}(t)+u_{rob}(t)$. In (a) and (b), the blue and red lines correspond to the result obtained with $u(t)=u_{nom}(t)+u_{rob}(t)$ and $u(t)=u_{nom}(t)$, respectively.}
			\label{fig:example1_time}
		\end{figure}
		
		\begin{figure}
			\centering
			\includegraphics[scale=1.2]{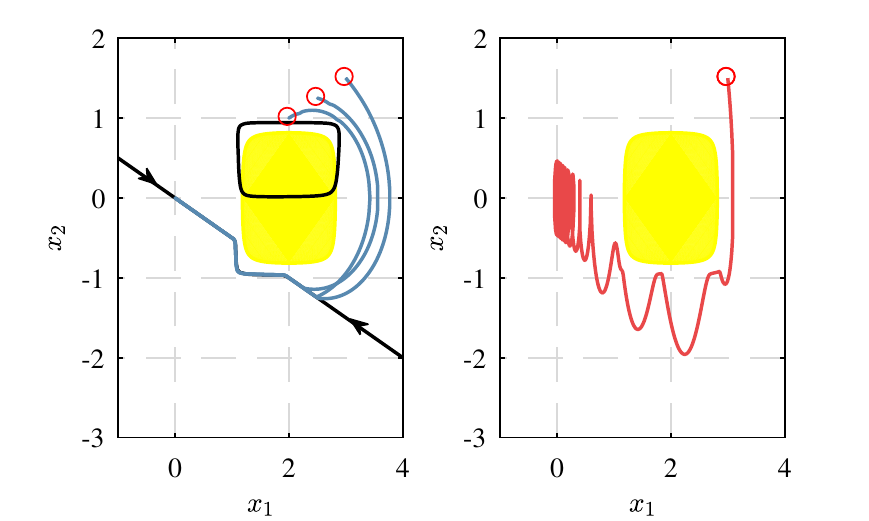}
			\caption{Left: Solution of the system in closed-loop with the robust controller. Right: Solution of the system in closed-loop with the nominal controller.}
			\label{fig:example1_states}
		\end{figure}
		
	\end{example}
	
\begin{figure}
		\centering
		\includegraphics[scale=1.2]{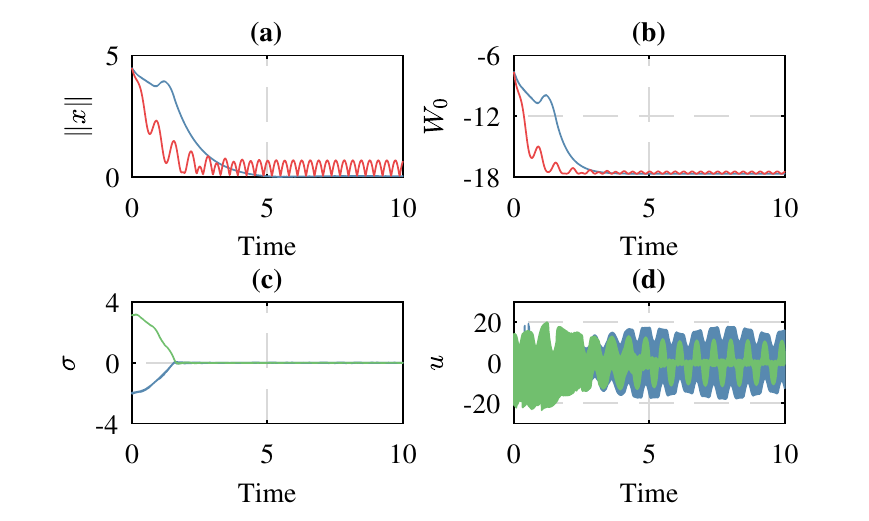}
		\caption{Results of the numerical simulation in Example \ref{ex:OngCortes}  with the initial condition $x_0=[2\:\: 2.5\:\: \pi]^T$. (a) Norm of the solution $x(t,x_0)$; (b) Function $t\mapsto W_0(x(t))$;  (c) Sliding variable $t\mapsto \sigma(t)$; (d) Control signal $u(t)=u_{nom}(t)+u_{rob}(t)$. In (a) and (b), the blue and red lines correspond to the result obtained with $u(t)=u_{nom}(t)+u_{rob}(t)$ and $u(t)=u_{nom}(t)$, respectively.}
		\label{fig:ex_ong_time}
	\end{figure}

\begin{figure}
		\centering
		\includegraphics[scale=1.2]{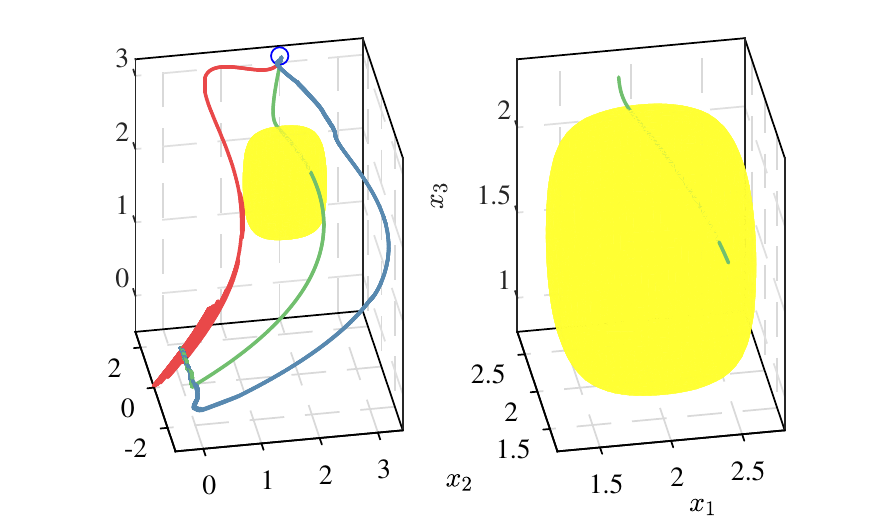}
		\caption{Left: Solution of the system in closed-loop with \eqref{ec:nominal_control} (red), \eqref{ec:robust_control} (blue) and \eqref{ec:robust_SMC}. Right: Zoom of the figure on the left.}
		\label{fig:ex_ong_states}
	\end{figure}

	\begin{example}
		\label{ex:OngCortes}
		We borrow the example presented in \cite{OngCortes2019universal} corresponding to a unicycle dynamics subjected to a drift given by a system of the form \eqref{ec:sys} with
		\begin{equation}
			f(x)=\begin{pmatrix}
				0\\ -x_2\\0
			\end{pmatrix}\:\:\text{and}\:\:g(x)=\begin{pmatrix}
				\cos (x_3) & 0\\
				\sin(x_3) & 0\\
				0 & 1
			\end{pmatrix}. 
		\end{equation}
		For simulation purposes,  we consider the disturbance and uncertain terms as
		\begin{equation*}
			\delta(t)=-5{\sin(10t)}\:\:\text{and}\:\:\Delta_g(t,x)=\begin{pmatrix}0.5\cos(x_3) & 0\\0 & 0\end{pmatrix}.
		\end{equation*}
		The unsafe set is
		\begin{equation*}
			\mathcal{D}=\left\{x\in\mathcal{X}:z(x)<4\right\},
		\end{equation*}
		where $\mathcal{X}=(1,3)\times (1,3)\times \left(\pi/2-1,\pi/2+1\right)$
		and
		\begin{equation*}
			z(x)=\dfrac{1}{1-(x_1-2)^2}+\dfrac{1}{1-(x_2-2)^2}+\dfrac{1}{1-(x_3-\pi/2)^2}.
		\end{equation*}
		As shown in \cite{OngCortes2019universal}, the function $V(x)=\dfrac{1}{2}\|x\|^2$ is a CLF. In order to construct a nominal control, we propose 
		\begin{equation*}
			\mathcal{B}(x)=\left\{\begin{array}{cc}
				e^{-z(x)}-e^{-4}, & \forall x\in \mathcal{X}\\
				-e^{-4}, & \text{otherwise}
			\end{array}\right..
		\end{equation*}
		Let us show that it is a CBF, that is, cf. \cite{RomdlonyJayawardhana2016}: (i) $ \mathcal{B}(x)>0\forall x\in \mathcal{D}$; (ii) $\nabla\mathcal{B}(x)\cdot g(x)= 0$ $\Rightarrow$ $\nabla\mathcal{B}(x)\cdot f(x)\leq 0$ for all $x\in \mathbb{R}^n\setminus \mathcal{D}$; and
		(iii) $\left\{x\in\mathbb{R}^n: \mathcal{B}(x)\leq 0\right\}\neq \emptyset$.
		
		The first and third conditions are immediate. Let us prove the second one.  Direct calculations lead to
		\begin{equation*}
			\nabla \mathcal{B}(x)\cdot g(x)=\begin{pmatrix}
				-\dfrac{\partial z}{\partial x_1}e^{-z}\cos(x_3)-\dfrac{\partial z}{\partial x_2}e^{-z(x)}\sin(x_3) & \dfrac{\partial z(x)}{\partial x_3}e^{-z(x)}
			\end{pmatrix}
		\end{equation*}
		for $x\in \mathcal{X}$. Then, it follows from $\nabla \mathcal{B}(x)\cdot g(x)=0$ that 
		\begin{equation*}
			\dfrac{\partial z(x)}{\partial x_3}=0\Leftrightarrow x_3=\pi/2
		\end{equation*}
		and
		\begin{equation*}
			\dfrac{\partial z(x)}{\partial x_2}=-\left. \dfrac{\cos(x_3)}{\sin(x_3)} \dfrac{\partial z(x)}{\partial x_1}\right|_{x_3=\pi/2}=0.
		\end{equation*}
		Thus, $\nabla \mathcal{B}(x)\cdot g(x)=0$ on $\{x\in \mathbb{R}^3:x_3=\pi/2\}$ and
		\begin{equation*}
			\nabla \mathcal{B}(x)\cdot f(x)=\dfrac{\partial z(x)}{\partial x_2}e^{-z(x)}x_2=0.
		\end{equation*}
		The constructed function $W_0$ is then 
		\begin{equation*}
			W_0(x)=\left\{\begin{array}{cc}
				\dfrac{1}{2}\|x\|^2+\lambda(e^{-z(x)}-e^{-4})+\kappa, & \forall x\in \mathcal{X}\\
				\dfrac{1}{2}\|x\|^2-\lambda e^{-4}+\kappa, & \text{otherwise}
			\end{array}\right.,
		\end{equation*} 
		with $\lambda=487.5263>(c_2c_4-c_1c_4)/e^{-4}$ and $\kappa=-c_1c_4=-8.7462$, where $c_1=1$, $c_2=2$, $c_3=\max_{x\in\partial \mathcal{X}}\|x\|^2=24.6$ and $c_4=\min_{\partial \mathcal{D}} \|x\|^2=8.74$.
		
		Notice that
		\begin{equation*}
			\nabla V(x)\cdot g(x)=\begin{pmatrix}
				\cos(x_3)x_1+\sin(x_3)x_2 & x_3
			\end{pmatrix}
		\end{equation*}
		so that
		\begin{equation*}
			\left\{x\in \mathbb{R}^3: \nabla V(x)\cdot g(x)=0\right\}=\left\{x\in\mathbb{R}^3:x_1=x_3=0, x_2\in \mathbb{R}\right\}\cap \mathcal{D}=\emptyset.
		\end{equation*}
		In view of the above, we can set $\mathcal{F}=\mathcal{X}$ and construct the sliding variable as 
		\begin{equation*}
			\sigma(t)=\nabla W_0(x)\cdot g(x)=\nabla V(x)\cdot g(x),\:x\in \mathbb{R}^n\setminus \mathcal{F}.
		\end{equation*}
		
		We consider the initial condition $x_0=[2\:\: 2.5\:\: \pi]^T$ and $\phi(t)=[-2\cos(t)\:\: \pi\cos(t)]$. It can be verified by direct calculation that $\sigma(0)=\phi(0)$ and that $t_f=\pi/2$. For the robust controller we set $q=2$, $\rho(t)=5$ for all $t\geq 0$ and $\mu=-0.5$. Figure 	\ref{fig:ex_ong_time} displays the results obtained for the norm of the states and the evolution of the energy function for both nominal and robust controllers, and the components of the sliding variable and the control signals of the robust controller.  Figure \ref{fig:ex_ong_states} illustrates the evolution of the trajectories in the space $(x_1,x_2,x_3)$ for three controllers: $u(t)=u_{nom}(t)$ in red, $u(t)=u_{nom}(t)+u_{rob}(t)$ in blue and  
		\begin{equation}
			\label{ec:robust_SMC}
			u(t)=u_{nom}(t)+k(t,x)\dfrac{ {\bar\sigma^T(t)}}{\|\bar \sigma(t)\|}
		\end{equation}
		in green. The difference between the second and third one relies on the incorporated transient function $\phi(t)$. The figure shows that the trajectory in green cross the unsafe set before reaching the sliding manifold. This case exemplifies what we previously discussed, namely, that without a suitable design of the reaching phase the closed-loop solution might be unsafe despite having a SSM.
		
	\end{example}
	
	\section{Conclusions and discussion}
	\label{sec:conclusion}
	
	We introduced the notion of safe sliding manifold, and describe its construction from the gradient of a Lyapunov-like energy function.  With the SSM at hand, 
	grounded on SMC theory and under some assumptions, we presented a controller for robust safety and stabilization of trajectories of systems with uncertainties and disturbances.  The proposed unit control ensures that the system trajectories safely converge to the SSM, within which there is an exact theoretical compensation of the uncertain and disturbance terms and the properties of the nominal design are recovered. 
	%Lyapunov-like energy function and a nominal control design are required
	
	{ An appealing and immediate solution to the  problem of robustifying a nominal design can be obtained by making safe and stable nominal solutions insensitive to uncertainties and disturbances from the initial time moment via integral sliding modes \cite{matthews1988decentralized, utkin1996integral,veselic2014sliding, veselic2015integral,rubagotti2011integral}. However,  being the nominal closed loop solution  the only feasible trajectory to avoid the unsafe set  is very restrictive, cf. with motion planning and obstacle avoidance algorithms.}
	
	Our proposal presents two main difficulties. First,  the knowledge of a {Lyapunov-like energy function and nominal control design are} required, and second, a transient time function  for a given initial condition is needed. Notice, however, that the presented approach is potentially useful  whenever suitable energy functions, {for instance, CLF and CBF}, are at hand, not only for the class of dynamical systems addressed in this paper, but also for others, e.g. time-delay systems.

	The proposed ideas bring to light several problems that might open several directions for future research.  Although  our starting point is a Lyapunov-like function for constructing a SSM, more work might be necessary to explore the design of SSM from the classical framework of SMC. Under this line the two issues that were stated within the introduction remain significant: how to construct/propose a SSM? how to ensure both finite time {convergence} and safety of the dynamics {to} the corresponding sliding manifold? In this regard, it is important to remark that using a manifold of relative degree one allows us to design a trajectory that avoids the unsafe set only before reaching the SSM.  Using higher order sliding mode controllers might be more complex in this setting, cf. \cite{levant2007integral}.

	%	\balance
	\bibliographystyle{IEEEtran}
	\bibliography{main_final}

\end{document}